\documentclass[superscriptaddress,aps,twocolumn,notitlepage,pra]{revtex4-2}
\usepackage{amsmath}
\usepackage{amssymb}
\usepackage{graphicx}
\usepackage{epstopdf}
\usepackage{mathtools}
\usepackage[colorlinks=true,linkcolor=blue,citecolor=blue]{hyperref}

\begin{document}

\title{Skin phase transition of a skyrmion in a ferromagnetic superfluid}

\author{Wooyoung Yun}
\affiliation{Department of Physics, Korea Advanced Institute of Science and Technology, Daejeon 34141, South Korea}

\author{Se Kwon Kim}
\email{sekwonkim@kaist.ac.kr}
\affiliation{Department of Physics, Korea Advanced Institute of Science and Technology, Daejeon 34141, South Korea}
\affiliation{Graduate School of Quantum Science and Technology, Korea Research Institute of Standards and Science, Daejeon 34141, Republic of Korea}

\author{Hiromitsu Takeuchi}
\email{takeuchi@omu.ac.jp}
\affiliation{Department of Physics, Osaka Metropolitan University, Osaka 558-8585, Japan}
\affiliation{Department of Physics, Nambu Yoichiro Institute of Theoretical and Experimental Physics (NITEP), Osaka 558-8585, Japan}

\begin{abstract}
    Compared with a scalar Bose-Einstein condensate (BEC), a spinor BEC supports richer phenomena because of its internal spin degrees of freedom, including topological solitons such as domain walls or skyrmions. Here, we theoretically reveal three distinct axisymmetric soliton structures formed as a circular domain wall in a spin-1 BEC, connected within fixed winding numbers. We refer to these structures as the broken-axisymmetry (BA)-skin, antiferromagnetic (AF)-skin and normal-core phases. Their occurrence is controlled by the two dimensionless parameters in the absence of linear Zeeman coefficient $p=0$: $q/\mu$, the ratio of the quadratic Zeeman coefficient to the chemical potential, and $c_s/c_n$, the ratio of spin interaction to the density interaction. The BA-skin and AF-skin phases exhibit local easy-plane magnetic order and local easy-plane spin-nematic order, respectively, while the normal-core phase is similar to a doubly quantized vortex analogous to that of a scalar BEC with a density depletion at the center. We obtain semi-analytical estimates of selected transition boundaries separating these phases by analyzing the Bogoliubov-de Gennes equation and also numerically confirm them by solving the Gross-Pitaevskii equation. We estimate the skyrmion radius using a hydrostatic approximation and compare the result with numerical solutions. We envision our results on skyrmions in a spin-1 BEC can trigger further investigation on novel soliton structures in a magnetic superfluid.
\end{abstract}
\maketitle

\section{Introduction}
A topological soliton is a nontrivial field configuration whose stability or classification is associated with the topology of the order parameter field. One of the most well-known examples is the skyrmion, which was proposed by Tony Skyrme as a field-theoretical model of baryons \cite{Skyrme1958, Skyrme1962}. This concept appears not only in nuclear physics, but also in various research fields such as magnetism \cite{Bogdanov1994,Lee2025,Speight2020,Okubo2012,Leonov2015}, liquid crystals \cite{Ivan2020,Zhao2023}, acoustics \cite{Ge2021} and superconductivity \cite{Speight2023,Benfenati2023}. Skyrmions are attracting attention in the physics community because their small size, robustness and controllability make them promising for data storage \cite{Fert2013,Nagaosa2013} and reservoir computing \cite{Beneke2024}.

One class of system that can host skyrmions is a spinor Bose-Einstein condensate (BEC) \cite{Zamora-Zamora2018,Mizushima2002,Mizushima2004,AlKhawaja2001,Isoshima2001,Takeuchi2022,Huh2025}. A Bose gas can undergo Bose-Einstein condensation, in which the lowest-energy single-particle state is macroscopically occupied below a critical temperature. BEC was first found in a single spin state of $^{87}$Rb \cite{Rb}, $^{23}$Na \cite{Na}, and $^{7}$Li \cite{Li}. In these systems, however, magnetic trapping confines only selected weak-field-seeking Zeeman states and thereby suppresses spinor dynamics, so the particles in these systems effectively behave as spinless bosons. Instead, one can realize BEC with an internal spin degree of freedom using an optical trap. The first realization of a spinor BEC used spin-1 $^{23}$Na \cite{Stamper1998} atoms in an optical trap. This BEC, which carries an internal spin degree of freedom, is called a spinor BEC and has become a platform for a broad range of research. For example, interparticle interaction can generate nontrivial spin textures and support topological solitons including skyrmions, domain walls \cite{Takeuchi2022}, hopfions \cite{Nitta2008}, monopoles \cite{Mikko2009}, and Alice rings \cite{Kivioja2023}.

In this work, we investigate skyrmions in spin-1 BEC. This system consists of spin-1 bosonic particles, and their spin degrees of freedom are represented by the Zeeman quantum number $m$. Therefore, the order parameter has three Zeeman components, $m=-1,0,+1$. In the parameter regime considered below, the system has a ferromagnetic ground state. Previous studies investigated skyrmions in the ferromagnetic phase of spin-1 BEC \cite{Zamora-Zamora2018,Mizushima2002,Mizushima2004,AlKhawaja2001,Isoshima2001}. However, it has been a common practice to neglect the quadratic Zeeman coefficient $q$, which can be controlled by the microwave dressing technique \cite{Gerbier2006,Bookjans2011,Zhao2014,Vinit2013,Jiang2014,Seo2015,Kim2021} and plays a role analogous to magnetic anisotropy. For example, the easy-axis phase is observed for negative $q$, whereas the easy-plane phase is observed for positive $q$ \cite{Kawaguchi2010}. The quadratic Zeeman effect has a nontrivial impact on the internal structure of topological solitons. One example is the effect of the quadratic Zeeman term on a flat domain wall in ferromagnetic spin-1 BEC \cite{Takeuchi2022}. The core of a domain wall exhibits a phase transition as the ratio $q/\mu$ is varied: If $0>q/\mu>-1$, the $m=0$ component has a nonzero amplitude, and the domain-wall core therefore forms the broken-axisymmetry (BA)-core with local easy-plane magnetic order. For $q/\mu<-1$, however, the $m=0$ component wave function vanishes, so the spin-1 BEC effectively reduces to a two-component BEC, and the domain wall core forms the antiferromagnetic (AF)-core with local easy-plane nematic-spin order. 

In this paper, we extend this research to integer skyrmions, which can be considered as circular domain walls in the ferromagnetic phase with easy-axis anisotropy for $q<0$ (F phase). We found that the internal structure of the skyrmion also exhibits the transitions both at its circular domain wall (skin) and its center similar to the flat domain wall. We theoretically predict the critical points of these phase transitions by analyzing the Bogoliubov theory and numerically confirm it by solving the Gross-Pitaevskii equation. We analytically estimate the radius for a representative $^7$Li condensate and compare the estimate with numerical results.

This paper is organized as follows. In Sec. \ref{section2}, we introduce the model of spin-1 BEC and skyrmion configuration. In Sec. \ref{section3}, we present the numerical solutions of skyrmion for $^7$Li condensate and analyze the skin phase transition between AF-skin phase and BA-skin phase using Bogoliubov-de Gennes equation. We also derive the skyrmion radius using the hydrostatic approximation. In Sec. \ref{section4}, we show the phase diagram for general spin-1 ferromagnetic BECs. The phase transition between AF-skin and Normal core phase is also described in this section. Finally, we summarize our results and make remarks on the stability of skyrmion and possibility of this phenomenon appearing in other systems in Sec. \ref{section5}.

\begin{figure*}[t]
    \centering
    \includegraphics[width=\textwidth]{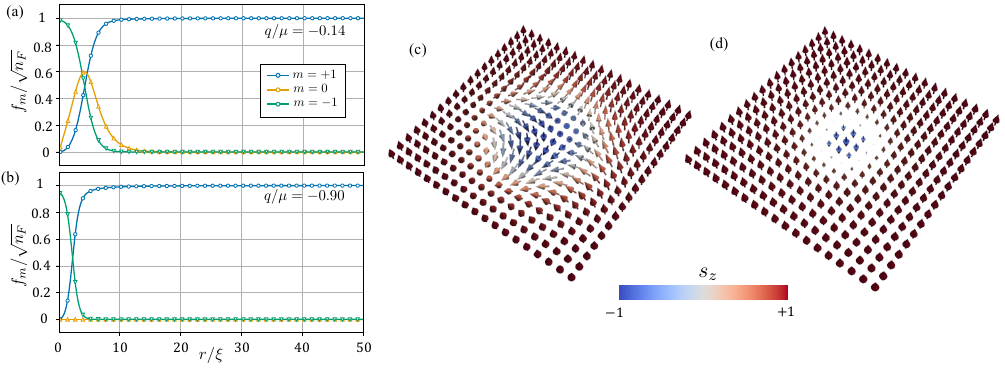}
    \caption{Plot of the order parameter amplitude $f_m(r)$ (solid lines) and the cross section of $|\Psi_m({\bf r})|$ (markers) for (a) $q/\mu=-0.14$ and (b) $q/\mu=-0.90$. Markers show $|\Psi_m(x, 0)|$ extracted from the two-dimensional simulation. In (a), the skyrmion has the BA-skin structure with nonzero $f_0$ at the domain wall. On the other hand, in (b), the skyrmion has the AF-skin structure with vanishing $f_0(r)$ and in-plane magnetization. (c-d) are the spin configuration for (c) $q/\mu=-0.14$ and (d) $q/\mu=-0.90$. Each arrow represents the spin direction, magnitude and the color represents the $z$-component of the spin $s_z$.}
    \label{fig1}
\end{figure*}

\section{Basic formulation}\label{section2}
In this section, we introduce the basic formulation used in the current work. 

\subsection{Gross-Pitaevskii equation}
Within the mean field theory, the dynamics of the spin-1 BEC is described by a three-component order parameter $\Psi({\bf r},t)=(\Psi_{+1}({\bf r}, t), \Psi_{0}({\bf r},t), \Psi_{-1}({\bf r},t))^T$ where $\Psi_m({\bf r},t)$ is the macroscopic wave function of the Zeeman component $m=-1,0,+1$. The order parameter dynamics is governed by the Gross-Pitaevskii equation (GPE), which can be derived from the following quasi-2D energy functional \cite{Kawaguchi2012,Takeuchi2022}:
\begin{align}
    E[\Psi]=\int &d^2x\bigg(\sum_m\frac{\hbar^2}{2M}\nabla\Psi_m^*\cdot\nabla\Psi_m+\frac{c_n}{2}n^2\nonumber\\
    &-\mu n+\frac{c_s}{2}{\bf s}^2-ps_z+q\Psi^\dagger\hat{\sigma}_z^2\Psi\bigg)~,\label{Energy_functional}
\end{align}
where $n$ is the particle density, ${\bf s}$ is the spin density, and $\hat{{\boldsymbol \sigma}}$ denotes the spin-1 matrices \cite{Kawaguchi2012}. These are given as
\begin{gather}
    n=\sum_m|\Psi_m|^2~,\\
    {\bf s}=(s_x,s_y,s_z)^T=\Psi^\dagger\hat{{\boldsymbol \sigma}}\Psi~,\\
    \hat{{\boldsymbol \sigma}}=(\hat{\sigma}_x,\hat{\sigma}_y,\hat{\sigma}_z)^T~.
\end{gather}
In Eq.~(\ref{Energy_functional}), $c_s$ and $c_n$ denote the spin- and density-dependent interaction coefficients, respectively. Here, $\mu$ is the chemical potential, while $p$ and $q$ are linear and quadratic Zeeman coefficients, respectively. We assume that $c_n$ and $\mu$ are positive such that the ground state becomes a uniform state with finite order parameter amplitudes. The chemical potential $\mu$ and linear Zeeman coefficient $p$ serve as the corresponding Lagrange multipliers associated with the conservation of the particle number and magnetization along the $z$-axis.

At $p=0$, a uniform spin-1 BEC exhibits four mean-field phases as functions of $q/\mu$ and $c_s/c_n$ \cite{Kawaguchi2012,Kawaguchi2010}. Throughout this paper, we omit a space-independent factor of the global phase associated with the broken U(1) symmetry, since it does not affect the physical properties of the system. When $c_s<0$ and $q<0$, the system is in the ferromagnetic (F) phase. The corresponding ground state wave function is given by $\Psi_{\pm1}=\sqrt{\frac{\mu-q\pm p}{c_n+c_s}}$, $\Psi_0=\Psi_{\mp1}=0$ and is doubly degenerate when $p=0$. This state is fully spin-polarized along the $z$-axis. If $c_s>0$ and $q<0$, the system is in the antiferromagnetic (AF) phase, whose ground state wave function satisfies $\Psi_0=0$ and $\Psi_{\pm1}\ne 0$. For $q>0$ and $c_s/c_n>-q/(2\mu)$, the system is in the polar (P) phase, characterized by $\Psi_{\pm1}=0$ and $\Psi_0\ne 0$. Finally, for $q>0$ and $c_s/c_n<-q/(2\mu)$, the system is in the broken-axisymmetry (BA) phase, in which all components $\Psi_m$ are nonzero. The condensate possesses both longitudinal and transverse magnetization, resulting in a spin polarization tilted away from the $z$-axis \cite{Murata2007,Kawaguchi2010}. In this paper, we restrict our parameters to $c_s<0$ and $q<0$ for the F phase.

\subsection{Vortex winding rule}
The GPE is derived from the functional derivative of Eq.~(\ref{Energy_functional}), $i\hbar\partial_t\Psi_m=\delta E/\delta\Psi_m^*$. For a stationary soliton state, we set $\Psi_m=\Phi_m$, $\partial_t\Phi_m=0$, the GPE is reduced to $\delta E/\delta\Psi_m^*=0$. In this work, we focus on skyrmions with rotational symmetry. We introduce the ansatz $\Phi_m(r,\theta)=f_m(r)e^{i\Theta_m}$ where $(r,\theta)$ are polar coordinates, $f_m(r)$ is real, and the phase $\Theta_m=L_m\theta+\varphi_m$. $L_m$ and $\varphi_m$ are the vortex number and constant phase of the $m$-component, respectively. We can obtain the reduced GPE by substituting the ansatz into Eq.~(\ref{Energy_functional}) and taking the functional derivative. These coupled equations are given by
\begin{align}
    0&=H_{\pm 1}f_{\pm 1}+c_sf_0^2(f_{\pm 1}+f_{\mp1}e^{-i\delta\Theta})~,\label{eq1_0}\\
    0&=H_0f_0+c_sf_0(f_{+1}^2+f_{-1}^2+2f_{+1}f_{-1}e^{i\delta\Theta})~,\label{eq2_0}
\end{align}
where
\begin{align}
    H_m&=-\frac{\hbar^2}{2M}\bigg(\frac{d^2}{dr^2}+\frac{1}{r}\frac{d}{dr}-\frac{L_m^2}{r^2}\bigg)\nonumber\\
    &\quad-\mu+c_nn+m^2q-mp+mc_ss_z~,\\
    \delta\Theta&=\Theta_{+1}+\Theta_{-1}-2\Theta_0\nonumber\\
    &=(L_{+1}+L_{-1}-2L_0)\theta+(\varphi_{+1}+\varphi_{-1}-2\varphi_0)~.
\end{align}
Here, $f_m(r)$ is a real-valued function and does not depend on $\theta$. When $c_s\ne 0$ and all of $f_{+1}$, $f_{-1}$ and $f_0$ are nonzero, $\delta\Theta=k\pi$ for an arbitrary integer $k$ for the radial equations to remain real. Therefore, the winding numbers of each component must satisfy the vortex winding rule \cite{Isoshima2001,Takeuchi2021,Takeuchi2021_2}:
\begin{gather}
    L_{+1}+L_{-1}-2L_0=0~.\label{winding_rule}
\end{gather}
We can set $\varphi_{+1}+\varphi_{-1}-2\varphi_0=0$ because with Eq.~(\ref{winding_rule}), the factor $e^{i\delta\Theta}=\pm 1$ may be absorbed into the real field $f_{+1}$ or $f_{-1}$. Finally, we have
\begin{align}
    0&=H_{\pm 1}f_{\pm 1}+c_sf_0^2(f_{\pm 1}+f_{\mp1})~,\label{eq1}\\
    0&=H_0f_0+c_sf_0(f_{+1}^2+f_{-1}^2+2f_{+1}f_{-1})~.\label{eq2}
\end{align}
Solving these equations yields the main results of this work. For a fixed set of $L_m$, after rescaling the length and order parameter amplitudes according to Eqs.~(\ref{eq1},\ref{eq2}),
\begin{gather}
    \xi=\frac{\hbar}{\sqrt{M(\mu-q)}},~n_F=\frac{\mu-q}{c_n+c_s}~, \label{rescale}
\end{gather}
the only remaining dimensionless parameters are $c_s/c_n$, $q/\mu$, and $p/\mu$. Here, $\xi$ and $n_F$ are the healing length and the ground-state density in the F phase for $p=0$, respectively.

\subsection{Skyrmion charge and vortex numbers}
Skyrmions are topological solitons characterized by topological charge. For a two-dimensional magnetic texture, a skyrmion is defined by the skyrmion charge $Q$ given as
\begin{gather}
    Q=-\frac{1}{4\pi}\int d^2x~{\bf m}\cdot(\partial_x{\bf m}\times\partial_y{\bf m})~,\label{skyrmion_charge}
\end{gather}
with normalized spin direction ${\bf m}=\mathbf{s}/|\mathbf{s}|$. In addition, without loss of generality, the spin points along $-\hat{z}$ at the center and approaches $+\hat{z}$ far from the skyrmion. For this skyrmion configuration, we need to set $L_{-1}=0$ because the $m=-1$ component occupies the center. Also, for a skyrmion with unit charge, we can set $L_{+1}=2$, equivalent to the vortex winding number $N_v$, using the Mermin-Ho relation $N_v=2Q$ \cite{Mermin1976,Takeuchi2022}. Therefore, vortex numbers of the three components are given as \cite{Zamora-Zamora2018,Ho1998}:
\begin{gather}
    L_{+1}=2,~L_0=1,~L_{-1}=0~.\label{skyrmion_winding}
\end{gather}
This relation also satisfies Eq.~(\ref{winding_rule}). With these vortex numbers, we can numerically solve Eqs.~(\ref{eq1},\ref{eq2}) and obtain the skyrmion configuration as shown in Fig.~\ref{fig1}. Since a skyrmion has a circular domain wall along its perimeter, we refer to the domain wall as the skin of the skyrmion. The skyrmion has a BA-skin structure with nonzero $f_0$ at the domain wall as shown in Fig.~\ref{fig1}.(a,c), whereas it has an AF-skin structure with vanishing $f_0$ and in-plane magnetization as shown in Fig.~\ref{fig1}.(b,d). Also, if only $m=+1$ component is nonzero, the skyrmion has a normal-core structure with a density depletion at the center, which is similar to a doubly quantized vortex in a scalar BEC. Note that BA-skin state and other states are topologically distinct because they have different skyrmion charges. The BA-skin state has $Q=1$, whereas the other states have $Q=0$, provided that the normalized spin field remains well defined.

Although fractional skyrmions have been observed experimentally \cite{Huh2025}, we focus on the integer skyrmions. In Ref.~\cite{Huh2025}, it was observed that integer skyrmions decompose into fractional skyrmions on a finite timescale. As can be seen from Eq.~(\ref{skyrmion_winding}), the decomposition of integer skyrmions is related to the splitting instability of quantum vortices with winding number 2. Since the splitting instability arises from the resonance between the phonons emitted during the splitting of the quantum vortex and the bound quasiparticles localized within the vortex \cite{Takeuchi2018}, the resonance conditions may not be satisfied depending on the wavenumber of the phonons or the system size, and the instability may not occur. In other words, integer skyrmions can have long lifetimes.

\section{\texorpdfstring{Skyrmion in ${}^{7}$Li condensates}{Skyrmion in 7Li condensates}}\label{section3}
We are particularly interested in skyrmions that could appear in the systems that undergo spontaneous breaking of the discrete symmetry associated with longitudinal magnetization \cite{Huh2020,Huh2024}. For simplicity, we restrict our analysis to $p=0$, which captures the essential physics while allowing for a simpler analytical treatment, and $c_s/c_n=-0.5$, which is a representative value for a $^7$Li condensate \cite{Huh2020}. The general case of $c_s/c_n$ will be discussed in the next section.

\subsection{BA-AF phase transition}
\begin{figure}[t]
    \centering
    \includegraphics[width=0.45\textwidth]{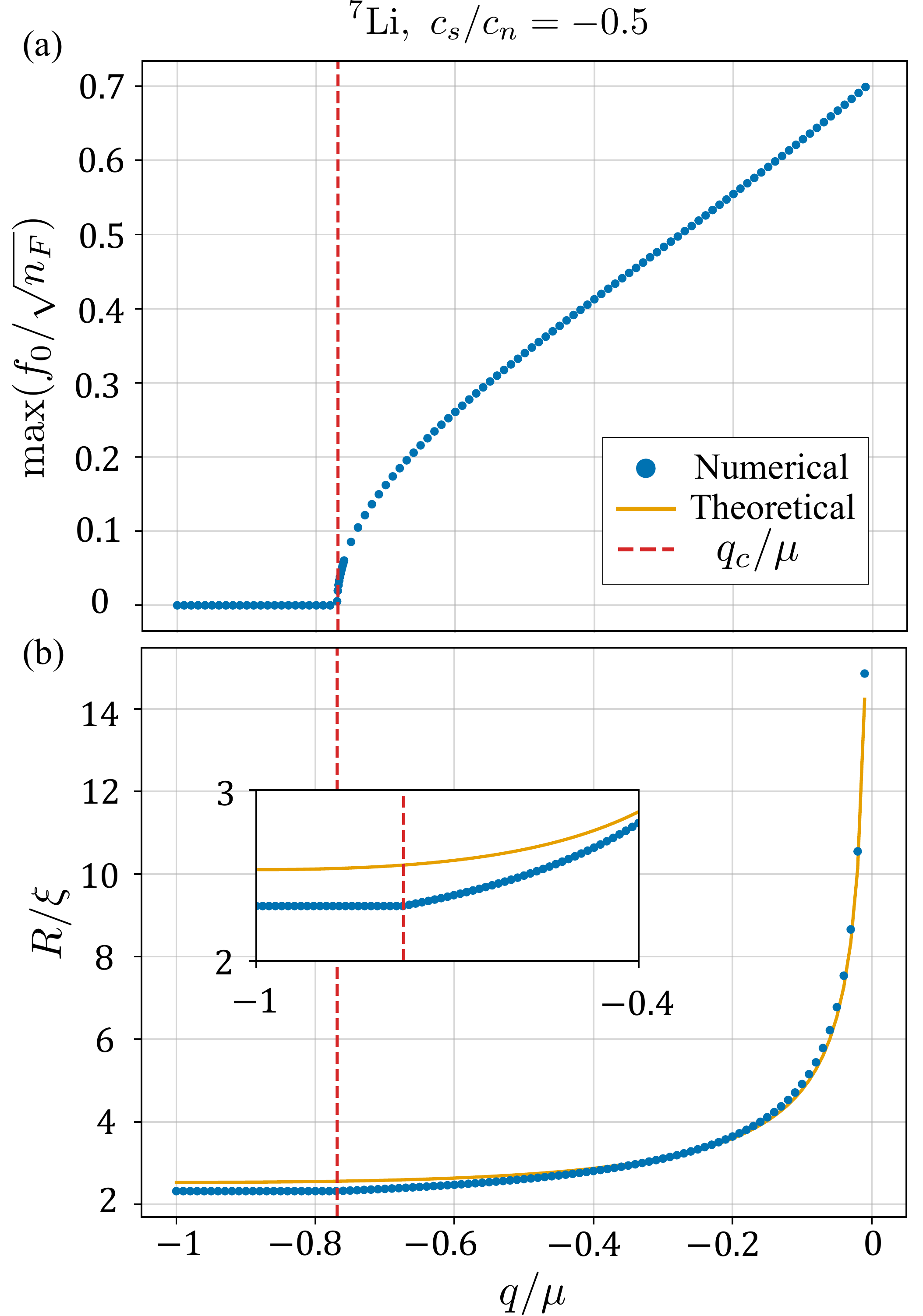}
    \caption{Numerical results for $^7$Li condensate. (a) The plot of $\max(f_0/\sqrt{n_F})$ as a function of $q/\mu$. The blue dots are the numerical values obtained by solving Eqs.~(\ref{eq1},\ref{eq2}) and the red dashed line is the critical value $q_c/\mu\approx-0.769$ obtained from BdG analysis. (b) The plots of skyrmion radius. The blue dots are the numerical values and the yellow solid line is the solution of Eq.~(\ref{radius_eq}), which is the theoretical prediction from hydrostatic approximation. The inset enlarges the vicinity of the critical point.}
    \label{fig_BA_AF}
\end{figure}

Using the vortex winding numbers in Eq.~(\ref{skyrmion_winding}), we solved the two-dimensional GPE and the corresponding three coupled radial GPEs [Eqs.~(\ref{eq1},\ref{eq2})] adjusting the parameter $q/\mu$. Representative solutions are shown in Fig.~\ref{fig_BA_AF}. In the numerical results, $f_0(r)$ has a nonzero value at the skyrmion boundary at smaller $|q/\mu|$, whereas it vanishes at larger $|q/\mu|$ and the system reduces to a two-component BEC. This is one of our main results: the skin structure of the skyrmion undergoes a BA-AF phase transition similar to the flat domain wall \cite{Takeuchi2022}. This phase transition is continuous as shown in Fig.~\ref{fig_BA_AF}.(a).

There is a critical value $q_c$ such that the axisymmetric solution has a BA skin for $q>q_c$ and an AF skin for $q<q_c$. We can qualitatively understand this phenomenon as the competition between the quadratic Zeeman term and the spin-dependent interaction. Since $q<0$ prefers the $m=\pm1$ components, the $m=0$ component is absent at larger $|q|$ and then the spin density vanishes inside the wall by forming the AF-skin. On the other hand, the spin-dependent interaction with $c_s<0$ prefers the BA-skin phase in the presence of the $m=0$ component at the domain wall to gain the ferromagnetic interaction energy with the transverse magnetization there. To determine the critical point $q_c$ precisely, we perform a linear-stability analysis based on the Bogoliubov-de Gennes (BdG) equations \cite{Takeuchi2022,Takeuchi2021}. We will present the details of this calculation in the following subsection.

\subsection{Two-mode approximation}
Since the $m=0$ component grows within the domain wall during the phase transition from the AF-skin to BA-skin, the transition is initiated by a fluctuation of $\Psi_0$. To analyze this transition, we introduce fluctuations $\delta\Psi_m$ around $\Phi_m$, the equilibrium state of the system. We write as $\Psi_m=\Phi_m+\delta\Psi_m$. Within Bogoliubov theory, these fluctuations are described as bosonic quasiparticles. For the AF-to-BA transition, we linearize about the AF-skin phase, for which $\Phi_0=0$.

To analytically elucidate the physical mechanism of phase transition, we introduce the perturbation approach. By linearizing the time-dependent GPE with respect to the fluctuations $\delta\Psi_m$, we obtain
\begin{align}
    i\hbar\partial_t\delta\Psi_{\pm1}&=-\frac{\hbar^2}{2M}\nabla^2\delta\Psi_{\pm1}-(\mu-q)\delta\Psi_{\pm1}\nonumber\\
    &+(c_n+c_s)(2|\Phi_{\pm1}|^2\delta\Psi_{\pm1}+\Phi_{\pm1}^2\delta\Psi_{\pm1}^*)\nonumber\\
    &+(c_n-c_s)(|\Phi_{\mp1}|^2\delta\Psi_{\pm1}+\Phi_{\mp1}^*\Phi_{\pm1}\delta\Psi_{\mp1}\nonumber\\
    &\qquad\qquad\quad+\Phi_{\mp1}\Phi_{\pm1}\delta\Psi_{\mp1}^*)~,\\
    i\hbar\partial_t\delta\Psi_0&=-\frac{\hbar^2}{2M}\nabla^2\delta\Psi_0-\mu\delta\Psi_0\nonumber\\
    &+2c_s\Phi_{+1}\Phi_{-1}\delta\Psi_0^*\nonumber\\
    &+(c_n+c_s)(|\Phi_{+1}|^2+|\Phi_{-1}|^2)\delta\Psi_0~.\label{flucm_0}
\end{align}
Here, we can analyze the dynamic instability by focusing on Eq.~(\ref{flucm_0}) only because this is decoupled from the other two equations for $\delta\Psi_{\pm1}$. To use the rotational symmetry, we substitute $f_m(r)e^{iL_m\theta}$ into $\Phi_m(r,\theta)$ and Eq.~(\ref{flucm_0}) reduces to
\begin{align}
    i\hbar\partial_t\delta\Psi_0&=-\frac{\hbar^2}{2M}\nabla^2\delta\Psi_0-\mu\delta\Psi_0\nonumber\\
    &+2c_sf_{+1}f_{-1}\delta\Psi_0^*e^{i(L_{+1}+L_{-1})\theta}\nonumber\\
    &+(c_n+c_s)(f_{+1}^2+f_{-1}^2)\delta\Psi_0~.\label{flucm_02}
\end{align}
To solve Eq.~(\ref{flucm_02}), we introduce the Bogoliubov transformation as
\begin{gather}
    \delta\Psi_0=e^{iL_0\theta}(u_0(r)e^{i(l\theta-\omega t)} - v_0^*(r)e^{-i(l\theta-\omega^* t)})~,\label{BdG_transform}
\end{gather}
with $L_0=1$ from Eq.~(\ref{skyrmion_winding}) and $l$ an integer denoting the angular momentum quantum number of the bosonic quasiparticle. Note that the frequency $\omega$ may be complex. If the imaginary part of $\omega$ is nonzero, the corresponding mode grows exponentially. This growth causes the dynamic instability and increases the population of the $m=0$ component. The appearance of such a mode signals the dynamic instability of the AF-skin phase and we can obtain $q_c$ from the condition that determines whether the imaginary part of $\omega$ is nonzero. By substituting Eq.~(\ref{BdG_transform}) into Eq.~(\ref{flucm_02}), the BdG equations for $u_0(r)$ and $v_0(r)$ are
\begin{align}
    \hbar\omega&\begin{pmatrix}
    u_0 \\ v_0
    \end{pmatrix}=\begin{pmatrix}
        M_+ & M_O \\
        -M_O^* & -M_-
    \end{pmatrix}\begin{pmatrix}
    u_0 \\ v_0
    \end{pmatrix}~,\label{BdG}
\end{align}
where
\begin{align}
    M_{\pm}&=-\frac{\hbar^2}{2M}\bigg(\frac{d^2}{dr^2}+\frac{1}{r}\frac{d}{dr}-\frac{(L_0\pm l)^2}{r^2}\bigg)-\mu\nonumber\\
    &+(c_n+c_s)(f_{+1}^2+f_{-1}^2)~,\\
    M_O&=-2c_sf_{+1}f_{-1}~,
\end{align}
To demonstrate that phase transition are caused by the coupling of these excitations, we can consider the perturbation expansion as follows:
\begin{gather}
    \hbar\omega\vec{u}=(\hat{h}_l+\delta\hat{h})\vec{u}~,\label{BdG_pert}
\end{gather}
where
\begin{gather}
    \hat{h}_l=\begin{pmatrix}
        h_l^+ & 0 \\
        0 & -h_l^-
    \end{pmatrix}~,\\
    \delta\hat{h}=\begin{pmatrix}
        c_s(f_{+1}^2+f_{-1}^2) & -2c_sf_{+1}f_{-1} \\
        2c_sf_{+1}f_{-1} & -c_s(f_{+1}^2+f_{-1}^2)
    \end{pmatrix}~,\label{perturbation}\\
    \vec{u}=(u_0,v_0)^T~,
\end{gather}
with
\begin{align}
    h_l^{\pm}&=-\frac{\hbar^2}{2M}\bigg(\frac{d^2}{dr^2}+\frac{1}{r}\frac{d}{dr}-\frac{(L_0\pm l)^2}{r^2}\bigg)\nonumber\\
    &-\mu+c_n(f_{+1}^2+f_{-1}^2)~.
\end{align}
Here, $L_0+l$ and $L_0-l$ are the vortex numbers of the fluctuations denoted by $u_0$ and $v_0$, respectively. The eigenvalue equation for the unperturbed Hamiltonian $\hat{h}_l$ is given as
\begin{gather}
    \hat{h}_l\vec{u}_+=\epsilon_+ \vec{u}_+,~\hat{h}_{-l}\vec{u}_-=\epsilon_- \vec{u}_-~,\label{unperturbed_eq}
\end{gather}
with eigenvectors $\vec{u}_+=(\phi(r),0)^T$, $\vec{u}_-=(0,\phi(r))^T$ and eigenvalues $\epsilon_+=-\epsilon_-$. These eigenvectors are normalized as
\begin{gather}
    \int d^2x~\phi^*(r)\phi(r)=1~,\\
    N_{\alpha\beta}=\int d^2x~\vec{u}^\dagger_\alpha\sigma_z\vec{u}_\beta=\pm\delta_{\alpha\beta}~,
\end{gather}
where $N_{\pm\pm}=\pm1$ and $\sigma_z=\text{diag}(1,-1)$. Equation~(\ref{unperturbed_eq}) reduces to
\begin{align*}
    \bigg[-\frac{1}{2}\bigg(\frac{d^2}{d\tilde{r}^2}+\frac{1}{\tilde{r}}\frac{d}{d\tilde{r}}-\frac{(L_0+l)^2}{\tilde{r}^2}\bigg)+V_\text{eff}\bigg]\phi&=\tilde{\epsilon}_+\phi~,
\end{align*}
where
\begin{gather*}
    V_\text{eff}=\frac{\tilde{f}_{+1}^2+\tilde{f}_{-1}^2}{1+c_s/c_n}~,
\end{gather*}
with $\tilde{r}=r/\xi$, $\tilde{f}_{\pm1}=f_{\pm1}/\sqrt{n_F}$, $\tilde{\epsilon}_\pm=(1\pm\epsilon_\pm/\mu)/(1-q/\mu)$. To obtain the critical value, we selected $l=0$ mode which gives $h_0^{+}=h_0^{-}=h_0$ and
\begin{align*}
    \bigg[-\frac{1}{2}\bigg(\frac{d^2}{d\tilde{r}^2}+\frac{1}{\tilde{r}}\frac{d}{d\tilde{r}}-\frac{1}{\tilde{r}^2}\bigg)+V_\text{eff}\bigg]\phi&=\tilde{\epsilon}_+\phi~.
\end{align*}
Numerical diagonalization gives
\begin{gather}
    \frac{\epsilon_\pm}{\mu-q}=\pm\bigg(\tilde{\epsilon}_\pm-\frac{1}{1-q/\mu}\bigg)~.\label{eigenvalue_unperturbed}
\end{gather}
To demonstrate that phase transitions are caused by the coupling of these excitations, we introduce the two-mode approximation as
\begin{gather}
    \vec{u}=C_+\vec{u}_+ + C_-\vec{u}_-~,\label{2modeapprox}
\end{gather}
Substituting Eq.~(\ref{2modeapprox}), projecting with $\vec{u}^\dagger_\pm\sigma_z$, and integrating over space yields
\begin{gather}
    \hbar\omega(C_+\vec{u}_+ + C_-\vec{u}_-)=(\hat{h}_0+\delta\hat{h})(C_+\vec{u}_+ + C_-\vec{u}_-)~,\nonumber\\
    \hbar\omega\begin{pmatrix}
    C_+ \\ C_-
    \end{pmatrix}=\begin{pmatrix}
    \epsilon_+ + M_{++} & M_{+-} \\
    M_{-+} & \epsilon_- + M_{--}
    \end{pmatrix}\begin{pmatrix}
    C_+ \\ C_-
    \end{pmatrix}~,\label{BdG_matrix}
\end{gather}
where
\begin{gather}
    M_{\alpha\beta}=N_{\alpha\alpha}\int d^2x\vec{u}_{\alpha}^\dagger\sigma_z\delta\hat{h}\vec{u}_{\beta}~.
\end{gather}
The solution of the eigenvalue equation Eq.~(\ref{BdG_matrix}) is given as
\begin{gather}
    \hbar\omega=\frac{e_++e_-}{2}\pm\sqrt{\bigg(\frac{e_+-e_-}{2}\bigg)^2+M_{+-}M_{-+}}~,
\end{gather}
with $e_\pm=\epsilon_\pm+M_{\pm\pm}$. The matrix elements $M_{\alpha\beta}$ are given by
\begin{align}
    M_{++}&=-M_{--}=\frac{c_s/c_n}{1+c_s/c_n}(\mu-q)I_1~,\label{M++}\\
    M_{+-}&=-M_{-+}=\frac{-2c_s/c_n}{1+c_s/c_n}(\mu-q)I_2~,\label{M+-}
\end{align}
where
\begin{align}
    I_1&=\int d^2x~(\tilde{f}_{+1}^2+\tilde{f}_{-1}^2)|\phi|^2~,\\
    I_2&=\int d^2x~\tilde{f}_{+1}\tilde{f}_{-1}\phi^2~.
\end{align}
The functions $\tilde{f}_{\pm 1}(r)$ are numerically obtained AF-skin phase solutions. Finally, we have
\begin{align}
    \hbar\omega&=\frac{\epsilon_++\epsilon_-}{2}\pm\sqrt{\bigg(\frac{\epsilon_+-\epsilon_-+2M_{++}}{2}\bigg)^2-M_{+-}^2}\nonumber\\
    &=\pm\sqrt{(\epsilon_++M_{++})^2-M_{+-}^2}\label{eigenvalue_BdG}~.
\end{align}
Dynamic instability occurs when $\mathrm{Im}(\omega)\ne 0$ since $\omega$ is a complex number, i.e., the term inside the square root in Eq.~(\ref{eigenvalue_BdG}) is negative. Therefore, the critical value $q_c$ is obtained by solving the following equation:
\begin{align}
    0=(\epsilon_++M_{++})^2-M_{+-}^2~.
\end{align}
Using Eqs.~(\ref{eigenvalue_unperturbed},\ref{M++},\ref{M+-}), we can rewrite this equation as
\begin{align}
    \frac{q_c}{\mu}&=1-\dfrac{1}{\tilde{\epsilon}_++\dfrac{c_s/c_n}{1+c_s/c_n}(I_1\pm 2I_2)}\label{qc}~.
\end{align}

For $q<0$, we selected the $I_1+2I_2$ to make $q_c<0$. In the case of $^7$Li condensate, $c_s/c_n=-0.5$, the numerical values are $I_1\approx0.563$, $I_2\approx0.142$, $\tilde{\epsilon}_+\approx1.412$, and $q_c/\mu\approx -0.769$. To confirm this value, we numerically solved 1-dimensional GPE [Eqs.~(\ref{eq1},\ref{eq2})] for various $q/\mu$ and plotted the maximum value of $f_0(r)$ as a function of $q/\mu$ in Fig.~\ref{fig_BA_AF}.(a).

\subsection{Skyrmion radius}
In this section, we estimate the skyrmion radius because it is the most apparent change induced by varying the parameter $q/\mu$. To quantify this radial response, we employed the hydrostatic approximation \cite{Hayashi2013,Takeuchi2021}, which is a low-energy effective theory formulated in terms of the relevant collective coordinates of the skyrmion. Higher-energy fluctuations around the skyrmion texture are neglected, yielding an effective energy $E_\text{eff}(R)$ for the skyrmion radius $R$. If the skyrmion radius is large enough compared to the skin width, each wave function component can be approximated by a step function, namely $f_{+1}=f_0=0$ for $r<R$ and $f_{-1}=f_0=0$ for $r>R$. By inserting these approximated functions into Eq.~(\ref{Energy_functional}) and neglecting the radial derivative terms, we obtained the effective energy as
\begin{align}
    E_\text{eff}(R)&=\int_0^{2\pi}d\theta\bigg(-\int_0^R rdr P_{-1} - \int_R^L rdr P_{+1} + \alpha R\bigg).\label{E_eff}
\end{align}
Here, $L$ is the system size, $P_{\pm1}(r)=(\tilde{\mu}(r)-\tilde{q}(r)\pm\tilde{p}(r))^2/2(c_n+c_s)$ is the pressure of each $m=\pm1$ component with the hydrostatic variables defined as
\begin{align}
    \tilde{\mu}(r)&=\mu-\frac{\hbar^2L_+^2}{2Mr^2}~,\\
    \tilde{p}(r)&=p-\frac{\hbar^2L_+L_-}{Mr^2}~,\\
    \tilde{q}(r)&=q+\frac{\hbar^2L_-^2}{2Mr^2}~,
\end{align}
with $L_{\pm}=(L_{+1}\pm L_{-1})/2$, and $\alpha$ is the tension coefficient of the curved domain wall. More specifically, by minimizing Eq.~(\ref{E_eff}) with respect to $R$, $\partial E_\text{eff}/\partial R=0$, we obtained the equilibrium skyrmion radius $R$ as a solution to the following equation:
\begin{gather}
    P_{-1}(R)-P_{+1}(R)=\frac{\alpha}{R}~.\label{pressure_balance}
\end{gather}
This equation is similar to the Young-Laplace equation in classical fluid mechanics \cite{LANDAU1987}. With the hydrostatic variables and Eq.~(\ref{skyrmion_winding}), Eq.~(\ref{pressure_balance}) reduces to
\begin{gather}
    2x^4-2x^2+\alpha'x=0~,\label{radius_eq}
\end{gather}
where $x=\xi/R$ and $\alpha=\alpha'(\mu-q)n_F\xi$. For radii much larger than the skin width, we approximated $\alpha$ by the tension of a flat domain wall. In this approximation, $\alpha'$ is a function of $q/\mu$ for $c_s/c_n=-0.5$ and the flat domain wall solution was obtained by numerical energy minimization (see Fig. 3 of Ref. \cite{Takeuchi2022}). We obtained the theoretical prediction of the skyrmion radius $R$ by solving Eq.~(\ref{radius_eq}) for various $q/\mu$. For $x>0$ this equation reduces to a cubic equation after excluding the $x=0$, or $R\to\infty$, solution. The solutions are given as
\begin{gather}
    x_k=\frac{2}{\sqrt{3}}\cos\bigg(\frac{\cos^{-1}(-3\sqrt{3}\alpha'/4)+2\pi k}{3}\bigg)~.\label{radis_eq_sol}\\
    (k=0,1,2)\nonumber
\end{gather}
Here, we can neglect $x_0$ and $x_1$ since $x_0$ corresponds to the local maximum of the effective energy and $x_1$ yields negative radius. Therefore, the stable solution is given as $R/\xi=1/x_2$.
We also numerically measured the radius of the skyrmion from the solutions of Eqs.~(\ref{eq1},\ref{eq2}) to confirm this prediction. The results are shown in Fig.~\ref{fig_BA_AF}.(b) as a function of $q/\mu$. One can see that the numerical results and the theoretical prediction from the hydrostatic approximation are in good agreement for large radius. However, there is a deviation for small radius in the vicinity of the critical point. This deviation is due to the curvature effect which is not included in the approximation. For large radii, the curvature is negligible so we can use $\alpha$ as the tension of the flat domain wall. On the other hand, for the small radius, the curvature is not negligible and the tension of the curved domain wall is different from that of the flat domain wall. This curvature effect can be included by considering the higher-order corrections in the hydrostatic approximation, but it is beyond the scope of the current work and we defer this extension to future work.

\begin{figure}[t]
    \centering
    \includegraphics[width=0.44\textwidth]{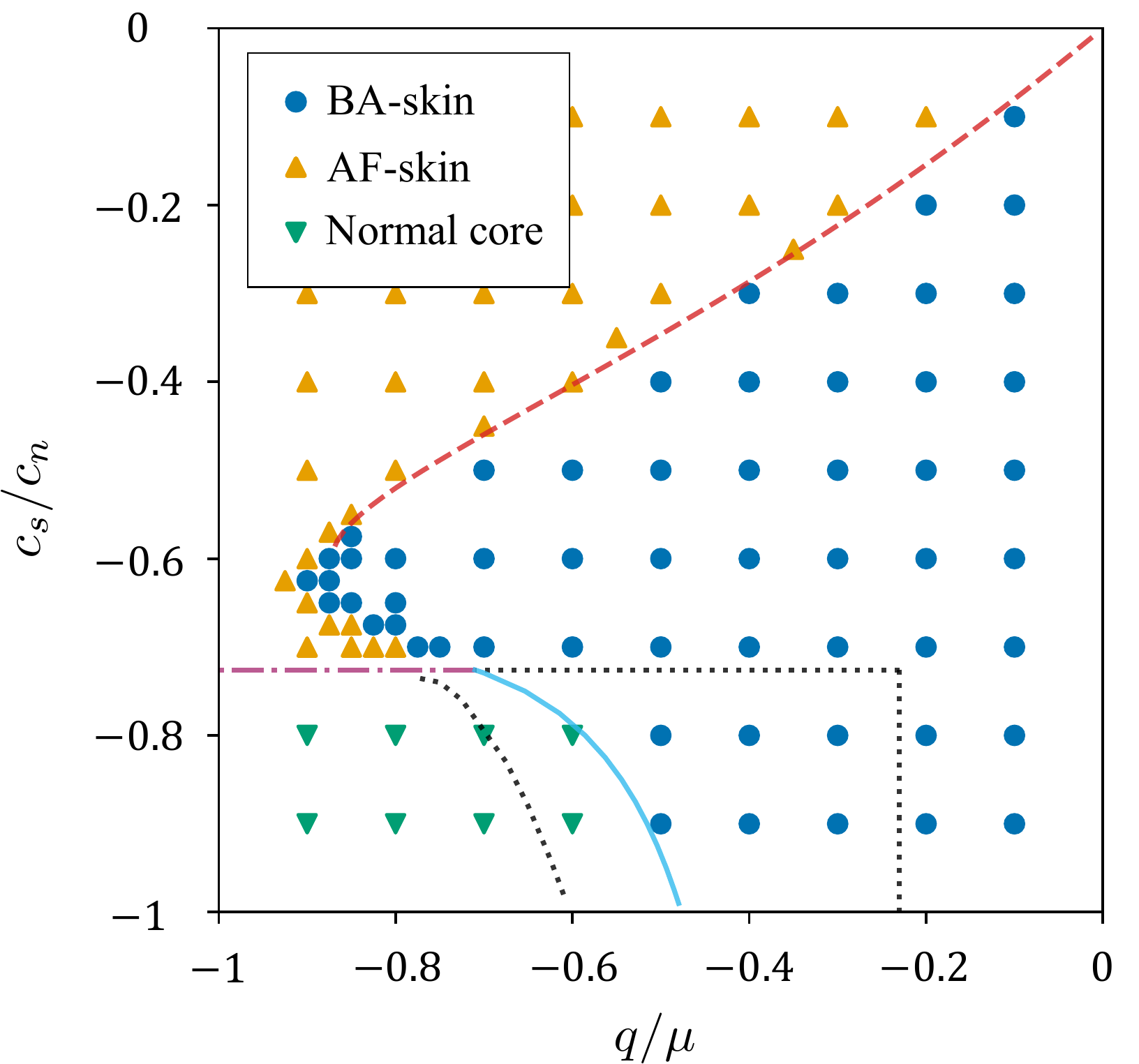}
    \caption{The phase diagram for $-1<c_s/c_n<0$ and $-1<q/\mu<0$ with $p=0$. The blue dots represent the BA-skin phase, the yellow triangles represent the AF-skin phase, and the green inverted triangles represent the normal-core vortex phase. The red dashed curve is the critical value for BA-AF transition, $q_c/\mu$ [Eq.~(\ref{qc})], the purple dash-dotted line is the critical value for AF-Nc transition [Eq.~(\ref{gc})]. The straight black dotted line parallel to $c_s/c_n$-axis is $q_{c1}$ [Eq.~(\ref{qc1})], parallel to $q/\mu$-axis is the $c_s/c_n=\gamma_c$ [Eq.~(\ref{gc2})]. The black dotted curve is the $q_{c2}$, which is obtained numerically by parameter scanning (see the discussion below Eq.~(\ref{gc2})). The cyan solid curve is the numerically-obtained coexistence curve for the discontinuous BA-Nc transition which satisfies $E[\Psi_\text{BA}]=E[\Psi_\text{Nc}]$.} 
    \label{fig_phase_diagram}
\end{figure}

\section{Phase diagram for the general case}\label{section4}
In the previous section, we investigated the skin phase transition of skyrmions at the fixed parameters $c_s/c_n=-0.5$ and $p=0$. Since these parameters can be tuned experimentally, it is important to examine how they affect the skyrmion phase transition. For example, $c_s/c_n$ can be controlled via Feshbach resonance \cite{Papoular2010} or by choosing different atomic species. Using numerical simulation and BdG analyses, we investigated the internal structure of a skyrmion and phase boundaries. 

In addition to the BA-skin and AF-skin phases, we found the normal-core (Nc) vortex phase, in which the $m=0$ and $m=-1$ components vanish while the $m=+1$ component forms a doubly quantized vortex, as in a scalar BEC. This phase is favored at large $|c_s/c_n|$ and large $|q/\mu|$. These three phases and their phase boundaries are summarized in the phase diagram in Fig.~\ref{fig_phase_diagram} and discussed in the following subsections.

\subsection{BA-AF phase transition}
We clearly observe a BA-skin-AF-skin transition analogous to the internal transition of a flat domain wall. Applying Eq.~(\ref{qc}) to general $c_s/c_n$ yields the red dashed curve in Fig.~\ref{fig_phase_diagram}. The numerical data agree well with this curve for $c_s/c_n\gtrsim -0.6$. Although not shown in the phase diagram, for $c_s/c_n\lesssim -0.6$, the BdG analysis still captures the overall trend, albeit with visible deviations. This could be because the perturbation term in Eq.~(\ref{BdG_pert}) becomes relatively large, making the perturbation approximation less accurate. Hence, the deviation increases with increasing $|c_s/c_n|$.

\subsection{AF-Nc phase transition}
The AF-Nc phase transition can be analyzed using a fluctuation analysis similar to the BA-AF phase transition. We consider fluctuations about the normal-core phase $\Phi_m$, writing $\Psi_m=\Phi_m+\delta\Psi_m$. Linearizing the time-dependent GPE with respect to $\delta\Psi_m$, we have
\begin{align}
    i\hbar\partial_t\delta\Psi_{+1}&=-\frac{\hbar^2}{2M}\nabla^2\delta\Psi_{+1}-(\mu-q)\delta\Psi_{+1}\nonumber\\
    &+(c_n+c_s)(2|\Phi_{+1}|^2\delta\Psi_{+1}+\Phi_{+1}^2\delta\Psi_{+1}^*)~,\label{flucm_m+1}\\
    i\hbar\partial_t\delta\Psi_{-1}&=-\frac{\hbar^2}{2M}\nabla^2\delta\Psi_{-1}-(\mu-q)\delta\Psi_{-1}\nonumber\\
    &+(c_n-c_s)|\Phi_{+1}|^2\delta\Psi_{-1}\label{flucm_m-1}~.
\end{align}

Equation~(\ref{flucm_m-1}) is decoupled from Eq.~(\ref{flucm_m+1}) and is already diagonal in the $\delta\Psi_{-1}$ sector. We solve it using the ansatz
\begin{gather}
    \delta\Psi_{-1}=u_{-1}(r)e^{iL_{-1}\theta}e^{i(l\theta-\omega t)}\label{u-1}~,\\
    \epsilon_{-1}u_{-1}=\bigg[-\frac{1}{2}\bigg(\frac{d^2}{d\tilde{r}^2}+\frac{1}{\tilde{r}}\frac{d}{d\tilde{r}}-\frac{(L_{-1}+l)^2}{\tilde{r}^2}\bigg)+U_\text{eff}\bigg]u_{-1}~,
\end{gather}
where
\begin{gather}
    U_\text{eff}=\frac{1-c_s/c_n}{1+c_s/c_n}\tilde{f}_{+1}^2,~\frac{\hbar\omega}{\mu-q}=\epsilon_{-1}-1~,
\end{gather}
with the effective potential $U_\text{eff}$ and the eigenvalue $\epsilon_{-1}$. As in the previous case, we need only consider the excitation at $l=0$. The metastability limit is given by
\begin{gather}
    \frac{c_s}{c_n}=\gamma_c\approx -0.726~,\quad\text{for } \frac{q}{\mu}\le \frac{q_\text{tri}}{\mu}~,\label{gc}
\end{gather}
with the triple point $q_\text{tri}/\mu\approx-0.762$, obtained by parameter scanning. The AF-Nc phase boundary [Eq.~(\ref{gc})] is depicted by the purple dash-dotted line in Fig.~\ref{fig_phase_diagram}. 

\subsection{BA-Nc phase transition}

\begin{figure}[t]
    \centering
    \includegraphics[width=0.48\textwidth]{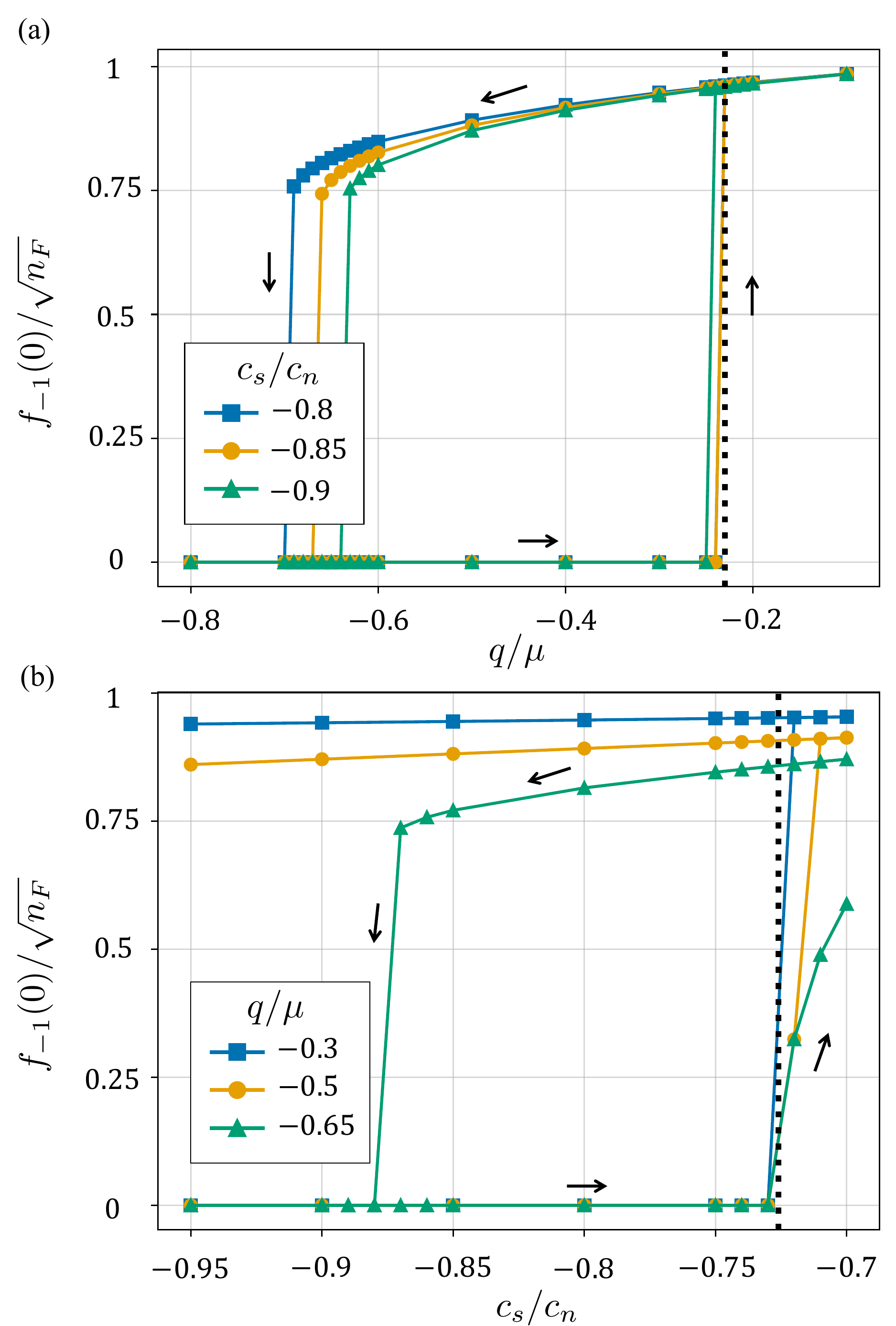}
    \caption{(a) The hysteresis of $f_{-1}(0)$ for $c_s/c_n=-0.8$, $-0.85$, and $-0.9$ as a function of $q/\mu$. The condition $f_{-1}(0)=0$ corresponds to the Nc phase, whereas $f_{-1}(0)\neq 0$ corresponds to the BA-skin phase. The black dotted line marks the spinodal $q_{c1}/\mu$ [Eq.~(\ref{qc1})]. (b) The hysteresis of $f_{-1}(0)$ for $q/\mu=-0.3$, $-0.5$, and $-0.65$ as a function of $c_s/c_n$. The black dotted line is the spinodal $\gamma_c$ [Eq.~(\ref{gc2})].}
    \label{fig_hysteresis}
\end{figure}

This phase transition is different from the previous two cases of BA-AF and AF-Nc phase transitions. Whereas a single order parameter serves as the instability mode in each of the preceding continuous transitions, this case requires considering two wave functions ($\Psi_0$ and $\Psi_{-1}$). Furthermore, in contrast to the preceding two continuous transitions, the present case exhibits a discontinuous phase transition, as evidenced by the hysteresis behavior in Fig.~\ref{fig_hysteresis}.

In this case, we consider fluctuations around the same normal-core vortex solution as in the analysis for AF-Nc transition. Then, we obtained the linearized GPE for $\delta\Psi_0$ as
\begin{align}
    i\hbar\partial_t\delta\Psi_0&=-\frac{\hbar^2}{2M}\nabla^2\delta\Psi_0-\mu\delta\Psi_0\nonumber\\
    &+(c_n+c_s)|\Phi_{+1}|^2\delta\Psi_0~.\label{flucm_0_BA-Nc}
\end{align}
This equation is decoupled from those for $\delta\Psi_{\pm 1}$ and already diagonalized for $\delta\Psi_0$. As in other cases, we need only consider the excitation at $l=0$. We used the ansatz $\delta\Psi_0=u_0(r)e^{iL_0\theta}e^{i(l\theta-\omega t)}$ which leads to the eigenvalue equation
\begin{gather}
    \epsilon_0u_0=\bigg[-\frac{1}{2}\bigg(\frac{d^2}{d\tilde{r}^2}+\frac{1}{\tilde{r}}\frac{d}{d\tilde{r}}-\frac{(L_0+l)^2}{\tilde{r}^2}\bigg)+\tilde{f}^2_{+1}\bigg]u_0~,\\
    \epsilon_0=\frac{\hbar\omega/\mu + 1}{1-q/\mu}~,
\end{gather}
with the effective potential $\tilde{f}^2_{+1}$ and the eigenvalue $\epsilon_0$. Since Landau instability occurs when $\omega<0$, the metastability limit is given by 
\begin{gather}
    \frac{q_{c1}}{\mu}=1-\frac{1}{\epsilon_0}\approx -0.230~,\quad\text{for }\frac{c_s}{c_n}\le\gamma_c\label{qc1}
\end{gather}
with the $\epsilon_0$ for the $l=0$ mode. This value is shown in Fig.~\ref{fig_phase_diagram} as the straight black dotted line parallel to $c_s/c_n$-axis. Since the BA-Nc transition is a discontinuous phase transition as shown in Fig.~\ref{fig_hysteresis}, $q_{c1}/\mu$ corresponds to a spinodal rather than the phase boundary. 

Another spinodal is obtained from the equation for $\delta\Psi_{-1}$, which is identical to Eq.~(\ref{flucm_m-1}) since the equilibrium state is the Nc phase in both cases. Therefore, the spinodal is given as
\begin{gather}
    \frac{c_s}{c_n}=\gamma_c\approx-0.726~,\quad\text{for }\frac{q_\text{tri}}{\mu}\le\frac{q}{\mu}\le \frac{q_{c1}}{\mu}~,\label{gc2}
\end{gather}
which is the same as Eq.~(\ref{gc}), although the applicable range is different. This spinodal is shown in Fig.~\ref{fig_phase_diagram} as a straight black dotted line parallel to $q/\mu$-axis.

The other spinodal, denoted by $q_{c2}$, should be calculated from the fluctuation equation around the equilibrium state of BA phase. However, the fluctuation problem about the BA phase does not reduce to a simple scalar eigenvalue equation. Therefore, we determined $q_{c2}$ numerically through parameter scanning and obtained the black dotted curve in Fig.~\ref{fig_phase_diagram}.

The critical curve of BA-Nc transition was determined by comparing the energies of the two phases, $E[\Psi_\text{BA}]$ and $E[\Psi_\text{Nc}]$, and was identified through numerical parameter scanning. It is shown as the cyan solid curve in Fig.~\ref{fig_phase_diagram}. We also numerically calculated the hysteresis of $f_{-1}(0)$ for various $c_s/c_n$ and $q/\mu$ to confirm the discontinuous phase transition and the spinodal points. The results are shown in Fig.~\ref{fig_hysteresis}(a) for fixed $c_s/c_n=-0.8$, $-0.85$, $-0.9$ and in Fig.~\ref{fig_hysteresis}(b) for fixed $q/\mu=-0.3$, $-0.5$, $-0.65$, respectively. The hysteresis loops clearly show the discontinuous phase transition, and the spinodal points are consistent with the values obtained from BdG analysis and parameter scanning.

\section{Summary and Discussion}\label{section5}
In this work, we investigated the skin phase transition of the skyrmion in a spin-1 BEC with a negative quadratic Zeeman coefficient. By numerically solving the Gross-Pitaevskii equation, we found that the skyrmion exhibits a continuous phase transition from the BA-skin to AF-skin analogous to that previously found for a flat domain wall in a $^7$Li condensate \cite{Takeuchi2022}. This phase transition is induced by the couling between the two modes with positive and negative norms of the $m=0$ component at the skyrmion radius. We also constructed the phase diagram for various spin-dependent interactions and quadratic Zeeman coefficients. Several boundaries were estimated semi-analytically using BdG analyses. In addition to the BA-skin and AF-skin phases, we found another phase: a normal-core (Nc) vortex phase similar to a doubly quantized vortex in a scalar BEC. The system also exhibits another continuous AF-Nc transition due to the Landau instability induced by excitations of the $m=-1$ component, while the discontinuous BA-Nc transition is characterized by the hysteresis of the $m=-1$ component. The critical points for these phase transitions were determined by the BdG analysis and numerical simulations.

We would like to make a few remarks on the stability mechanism for skyrmions in our system. In 2-dimensional chiral magnets, the skyrmion structure is stabilized by Dzyaloshinskii-Moriya interaction (DMI) \cite{Bogdanov1994,Lee2025} or magnetic frustration \cite{Okubo2012,Leonov2015,Speight2020}, whereas Eq.~(\ref{Energy_functional}) does not have such terms. However, stability arises from Eq.~(\ref{skyrmion_winding}), which fixes the vortex winding numbers for each component. $L_{+1}$ and $L_0$ are nonzero so that the densities of the $m=0$ and $m=+1$ components show depletions at the skyrmion core and the $m=-1$ component occupies the core region because density depletion is energetically unfavorable. This structure can produce a finite-radius stationary skyrmion within the fixed winding numbers [Eq.~(\ref{skyrmion_winding})]. In addition, spin winding also follows from these winding numbers. The $x$ and $y$ components of spin density are
\begin{align*}
    s_x&=\sqrt{2}(f_{+1}f_0\cos((L_{+1}-L_0)\theta) \\
    &\qquad+ f_{-1}f_0\cos((L_{-1}-L_0)\theta))~,\\
    &=\sqrt{2}f_0(f_{+1}+f_{-1})\cos\theta~,\\
    s_y&=\sqrt{2}(f_{+1}f_0\sin((L_0-L_{+1})\theta) \\
    &\qquad - f_{-1}f_0\sin((L_0-L_{-1})\theta))~,\\
    &=-\sqrt{2}f_0(f_{+1} + f_{-1})\sin\theta~.
\end{align*}
This result indicates that the in-plane spin density forms a vortex-like structure with unit-magnitude winding number around the skyrmion center. Therefore, a spinor BEC can host an axisymmetric spin-vortex texture without DMI or magnetic frustration.

In low-temperature two-dimensional chiral magnets, skyrmions are usually studied within the standard micromagnetic approximation in which the local magnetization is constrained to its saturation value, i.e., the normalized order parameter satisfies $|{\bf s}|=1$ everywhere. Such fixed-length micromagnetic models do not support the amplitude-vanishing transition found here in a spinor BEC. However, we expect that ferroelectric systems can exhibit a qualitatively related phenomenon, although the spin-nematic order of the present AF skin has no direct counterpart in a polarization-only theory. This is because the order parameter of these systems is electric polarization ${\bf P}$ and its magnitude may vary spatially, as does the spin amplitude in a spinor BEC. In these systems, AF-core domain walls are called Ising-type domain walls, which have been studied \cite{Gu2014,LeeDH2009}.

\section*{Acknowledgments}
W.Y. and S.K.K. were supported by Brain Pool Plus Program through the National Research Foundation of Korea funded by the Ministry of Science and ICT (2020H1D3A2A03099291), Basic Science Research Program through the National Research Foundation of Korea (NRF) funded by the Ministry of Education (2019R1A6A1A10073887), and Creation of the quantum information science R\&D ecosystem (based on human resources) through the National Research Foundation of Korea (NRF) funded by the Korean government (Ministry of Science and ICT (MSIT)) (RS-2023-00256050). HT is supported by JSPS KAKENHI Grants No. JP18KK0391, No. JP20H01842, and No. JP26K00640; and JST, PRESTO (Japan) Grant No. JPMJPR23O5.

\section*{Appendix A: Numerical method}
In this section, we explain the numerical method used in this research. First, to obtain the equilibrium state solution of Eqs.~(\ref{eq1},\ref{eq2}), we used the imaginary-time evolution method. We used the finite-difference method for spatial discretization and numerical differentiation. The spatial coordinate is discretized as $r\approx r_i=h+ih~(i=0,\cdots,499)$ where $h=0.1\xi$ is the spatial step size and the total system size is $L=50\xi$, which is sufficiently large compared to the skyrmion size. The numerical derivatives, $df_m/dr$ and $d^2f_m/dr^2$, are computed by second-order central differences for both the first and second derivatives, respectively. The boundary conditions are as follows: at the system boundary, $r=L$, the Neumann boundary condition is imposed as $df_m/dr|_{r=L}=0$ for every component. At the skyrmion center, $r=0$, the boundary conditions were imposed according to the regularity of the radial wave functions. Since the vortex winding numbers are given by Eq.~(\ref{skyrmion_winding}), the radial wave functions behave as $f_m(r)\sim r^{|L_m|}$ near the origin. Accordingly, we imposed $f_{+1}(0)=f_0(0)=0$ and $df_{-1}/dr|_{r=0}=0$. Similar boundary conditions are applied to the numerical diagonalization of BdG equations. The initial state is set to the skyrmion ansatz and the imaginary time step is set to $\Delta t=0.0025\hbar/(\mu-q)$.

We solved a 2-dimensional GPE similar to the 1-dimensional case. The spatial coordinate is discretized as $(x,y)\approx(x_i,y_j)=(-L+ih,-L+jh)$ where $h=0.25\xi$ and $i,j=0,\cdots,400$. The Neumann boundary condition is imposed at the system boundary. All GPE simulations are implemented by CUDA C++ and performed using dimensionless variables. Lengths are expressed in units of $\xi$, and order parameter amplitudes in units of $\sqrt{n_F}$ variables.

The skyrmion radius was measured from the numerical solution of Eqs.~(\ref{eq1},\ref{eq2}). We defined the skyrmion radius $R$ as the radius where $s_z(R)=0$. To measure this point, we first calculated $s_z(r_i)$ at each spatial grid point $r_i$. Then, we identified adjacent points $r_{k}$ and $r_{k+1}$ that satisfy $s_z(r_{k})$ is the maximum value among the points $s_z<0$ and $s_z(r_{k+1})$ is the minimum value among the points $s_z>0$. After that, we performed linear interpolation between these two points and obtained the skyrmion radius $R$.

\section*{Appendix B: Solution of the radius equation}
In this section, we explain the analytic solution of Eq.~(\ref{radius_eq}). This equation can be factored as
\begin{gather}
    x(2x^3 - 2x + \alpha')=0~.\label{radius_eq_factor}
\end{gather}
The cubic equation in parentheses can be solved by substituting $x=2\cos\theta/\sqrt{3}$. The cubic equation becomes
\begin{align}
    4\cos^3\theta - 3\cos\theta + \frac{3\sqrt{3}\alpha'}{4} &= 0~.\label{cubic_eq}
\end{align}
Using the triple-angle formula for cosine, Eq.~(\ref{cubic_eq}) reduces to
\begin{gather*}
    \cos 3\theta = 4\cos^3\theta - 3\cos\theta = -\frac{3\sqrt{3}\alpha'}{4}~.
\end{gather*}
Since cosine is a $2\pi$-periodic function, the general solution is given as
\begin{align*}
    3\theta &= \cos^{-1}\bigg(-\frac{3\sqrt{3}\alpha'}{4}\bigg)+2\pi k,~(k\in\mathbb{Z})\\
    \theta &= \frac{1}{3}\cos^{-1}\bigg(-\frac{3\sqrt{3}\alpha'}{4}\bigg)+\frac{2\pi k}{3}~.
\end{align*}
The distinct three roots are therefore
\begin{gather*}
    x_k=\frac{2}{\sqrt{3}}\cos\bigg(\frac{\cos^{-1}(-3\sqrt{3}\alpha'/4)+2\pi k}{3}\bigg)~.\\
    (k=0,1,2)
\end{gather*}

\bibliography{ref.bib}

@article{Skyrme1958,
author = {Skyrme, T. H. R. },
title = {A non-linear theory of strong interactions},
journal = {Proceedings of the Royal Society of London. Series A. Mathematical and Physical Sciences},
volume = {247},
number = {1249},
pages = {260-278},
year = {1958},
doi = {10.1098/rspa.1958.0183},
URL = {https://royalsocietypublishing.org/doi/abs/10.1098/rspa.1958.0183}
}

@article{Skyrme1962,
title = {A unified field theory of mesons and baryons},
journal = {Nuclear Physics},
volume = {31},
pages = {556-569},
year = {1962},
issn = {0029-5582},
doi = {https://doi.org/10.1016/0029-5582(62)90775-7},
url = {https://www.sciencedirect.com/science/article/pii/0029558262907757},
author = {T.H.R. Skyrme}
}

@article{Stamper1998,
  title = {Optical Confinement of a Bose-Einstein Condensate},
  author = {Stamper-Kurn, D. M. and Andrews, M. R. and Chikkatur, A. P. and Inouye, S. and Miesner, H.-J. and Stenger, J. and Ketterle, W.},
  journal = {Phys. Rev. Lett.},
  volume = {80},
  issue = {10},
  pages = {2027--2030},
  numpages = {0},
  year = {1998},
  month = {Mar},
  publisher = {American Physical Society},
  doi = {10.1103/PhysRevLett.80.2027},
  url = {https://link.aps.org/doi/10.1103/PhysRevLett.80.2027}
}

@article{Rb,
author = {M. H. Anderson  and J. R. Ensher  and M. R. Matthews  and C. E. Wieman  and E. A. Cornell },
title = {Observation of Bose-Einstein Condensation in a Dilute Atomic Vapor},
journal = {Science},
volume = {269},
number = {5221},
pages = {198-201},
year = {1995},
doi = {10.1126/science.269.5221.198},
URL = {https://www.science.org/doi/abs/10.1126/science.269.5221.198}}

@article{Na,
  title = {Bose-Einstein Condensation in a Gas of Sodium Atoms},
  author = {Davis, K. B. and Mewes, M. -O. and Andrews, M. R. and van Druten, N. J. and Durfee, D. S. and Kurn, D. M. and Ketterle, W.},
  journal = {Phys. Rev. Lett.},
  volume = {75},
  issue = {22},
  pages = {3969--3973},
  numpages = {0},
  year = {1995},
  month = {Nov},
  publisher = {American Physical Society},
  doi = {10.1103/PhysRevLett.75.3969},
  url = {https://link.aps.org/doi/10.1103/PhysRevLett.75.3969}
}

@article{Li,
  title = {Bose-Einstein Condensation of Lithium: Observation of Limited Condensate Number},
  author = {Bradley, C. C. and Sackett, C. A. and Hulet, R. G.},
  journal = {Phys. Rev. Lett.},
  volume = {78},
  issue = {6},
  pages = {985--989},
  numpages = {0},
  year = {1997},
  month = {Feb},
  publisher = {American Physical Society},
  doi = {10.1103/PhysRevLett.78.985},
  url = {https://link.aps.org/doi/10.1103/PhysRevLett.78.985}
}

@article{Zamora-Zamora2018,
title = {Skyrmions with arbitrary topological charges in spinor Bose–Einstein condensates},
doi = {10.1088/1361-6455/aaa324},
url = {https://doi.org/10.1088/1361-6455/aaa324},
year = {2018},
month = {jan},
publisher = {IOP Publishing},
volume = {51},
number = {4},
pages = {045301},
author = {Zamora-Zamora, R and Romero-Rochín, V},
journal = {Journal of Physics B: Atomic, Molecular and Optical Physics}
}

@article{Takeuchi2021,
  title = {Phase diagram of vortices in the polar phase of spin-1 Bose-Einstein condensates},
  author = {Takeuchi, Hiromitsu},
  journal = {Phys. Rev. A},
  volume = {104},
  issue = {1},
  pages = {013316},
  numpages = {15},
  year = {2021},
  month = {Jul},
  publisher = {American Physical Society},
  doi = {10.1103/PhysRevA.104.013316},
  url = {https://link.aps.org/doi/10.1103/PhysRevA.104.013316}
}

@article{Takeuchi2022,
	author = {Takeuchi, Hiromitsu},
	doi = {10.1103/PhysRevA.105.013328},
	issue = {1},
	journal = {Phys. Rev. A},
	month = {Jan},
	numpages = {14},
	pages = {013328},
	publisher = {American Physical Society},
	title = {Spin-current instability at a magnetic domain wall in a ferromagnetic superfluid: A generation mechanism of eccentric fractional skyrmions},
	url = {https://link.aps.org/doi/10.1103/PhysRevA.105.013328},
	volume = {105},
	year = {2022}}

@article{Kawaguchi2012,
title = {Spinor {Bose–Einstein condensates}},
journal = {Phys. Rep.},
volume = {520},
number = {5},
pages = {253-381},
year = {2012},
issn = {0370-1573},
doi = {https://doi.org/10.1016/j.physrep.2012.07.005},
url = {https://www.sciencedirect.com/science/article/pii/S0370157312002098},
author = {Yuki Kawaguchi and Masahito Ueda}
}

@article{Kawaguchi2010,
  title = {Spontaneous magnetic ordering in a ferromagnetic spinor dipolar {Bose-Einstein condensate}},
  author = {Kawaguchi, Yuki and Saito, Hiroki and Kudo, Kazue and Ueda, Masahito},
  journal = {Phys. Rev. A},
  volume = {82},
  issue = {4},
  pages = {043627},
  numpages = {17},
  year = {2010},
  month = {Oct},
  publisher = {American Physical Society},
  doi = {10.1103/PhysRevA.82.043627},
  url = {https://link.aps.org/doi/10.1103/PhysRevA.82.043627}
}

@article{Murata2007,
  title = {Broken-axisymmetry phase of a spin-1 ferromagnetic Bose-Einstein condensate},
  author = {Murata, Keiji and Saito, Hiroki and Ueda, Masahito},
  journal = {Phys. Rev. A},
  volume = {75},
  issue = {1},
  pages = {013607},
  numpages = {10},
  year = {2007},
  month = {Jan},
  publisher = {American Physical Society},
  doi = {10.1103/PhysRevA.75.013607},
  url = {https://link.aps.org/doi/10.1103/PhysRevA.75.013607}
}

@article{Ho1998,
  title = {Spinor Bose Condensates in Optical Traps},
  author = {Ho, Tin-Lun},
  journal = {Phys. Rev. Lett.},
  volume = {81},
  issue = {4},
  pages = {742--745},
  numpages = {0},
  year = {1998},
  month = {Jul},
  publisher = {American Physical Society},
  doi = {10.1103/PhysRevLett.81.742},
  url = {https://link.aps.org/doi/10.1103/PhysRevLett.81.742}
}

@article{Hayashi2013,
  title = {Instability crossover of helical shear flow in segregated Bose-Einstein condensates},
  author = {Hayashi, Shinsuke and Tsubota, Makoto and Takeuchi, Hiromitsu},
  journal = {Phys. Rev. A},
  volume = {87},
  issue = {6},
  pages = {063628},
  numpages = {9},
  year = {2013},
  month = {Jun},
  publisher = {American Physical Society},
  doi = {10.1103/PhysRevA.87.063628},
  url = {https://link.aps.org/doi/10.1103/PhysRevA.87.063628}
}

@article{Isoshima2001,
author = {Isoshima ,Tomoya and Machida ,Kazushige and Ohmi ,Tetsuo},
title = {Quantum Vortex in a Spinor Bose-Einstein Condensate},
journal = {Journal of the Physical Society of Japan},
volume = {70},
number = {6},
pages = {1604-1610},
year = {2001},
doi = {10.1143/JPSJ.70.1604},
url = {https://doi.org/10.1143/JPSJ.70.1604}
}

@Article{AlKhawaja2001,
author={Al Khawaja, Usama
and Stoof, Henk},
title={Skyrmions in a ferromagnetic Bose--Einstein condensate},
journal={Nature},
year={2001},
month={Jun},
day={01},
volume={411},
number={6840},
pages={918-920},
issn={1476-4687},
doi={10.1038/35082010},
url={https://doi.org/10.1038/35082010}
}

@article{Nitta2008,
  title = {Knots in a Spinor {Bose-Einstein Condensate}},
  author = {Kawaguchi, Yuki and Nitta, Muneto and Ueda, Masahito},
  journal = {Phys. Rev. Lett.},
  volume = {100},
  issue = {18},
  pages = {180403},
  numpages = {4},
  year = {2008},
  month = {May},
  publisher = {American Physical Society},
  doi = {10.1103/PhysRevLett.100.180403},
  url = {https://link.aps.org/doi/10.1103/PhysRevLett.100.180403}
}

@article{Mikko2009,
  title = {Creation of {Dirac} Monopoles in Spinor Bose-Einstein Condensates},
  author = {Pietil\"a, Ville and M\"ott\"onen, Mikko},
  journal = {Phys. Rev. Lett.},
  volume = {103},
  issue = {3},
  pages = {030401},
  numpages = {4},
  year = {2009},
  month = {Jul},
  publisher = {American Physical Society},
  doi = {10.1103/PhysRevLett.103.030401},
  url = {https://link.aps.org/doi/10.1103/PhysRevLett.103.030401}
}

@article{Kivioja2023,
  title = {Evolution and decay of an {Alice} ring in a spinor Bose-Einstein condensate},
  author = {Kivioja, Markus and Zamora-Zamora, Roberto and Blinova, Alina and M\"onk\"ol\"a, Sanna and Rossi, Tuomo and M\"ott\"onen, Mikko},
  journal = {Phys. Rev. Res.},
  volume = {5},
  issue = {2},
  pages = {023104},
  numpages = {10},
  year = {2023},
  month = {May},
  publisher = {American Physical Society},
  doi = {10.1103/PhysRevResearch.5.023104},
  url = {https://link.aps.org/doi/10.1103/PhysRevResearch.5.023104}
}

@article{Mermin1976,
  title = {Circulation and Angular Momentum in the $A$ Phase of Superfluid Helium-3},
  author = {Mermin, N. D. and Ho, Tin-Lun},
  journal = {Phys. Rev. Lett.},
  volume = {36},
  issue = {11},
  pages = {594--597},
  numpages = {0},
  year = {1976},
  month = {Mar},
  publisher = {American Physical Society},
  doi = {10.1103/PhysRevLett.36.594},
  url = {https://link.aps.org/doi/10.1103/PhysRevLett.36.594}
}

@article{Papoular2010,
  title = {Microwave-induced {Fano-Feshbach} resonances},
  author = {Papoular, D. J. and Shlyapnikov, G. V. and Dalibard, J.},
  journal = {Phys. Rev. A},
  volume = {81},
  issue = {4},
  pages = {041603},
  numpages = {4},
  year = {2010},
  month = {Apr},
  publisher = {American Physical Society},
  doi = {10.1103/PhysRevA.81.041603},
  url = {https://link.aps.org/doi/10.1103/PhysRevA.81.041603}
}

@article{Lee2025,
  title = {Domain wall networks as skyrmion crystals in chiral magnets},
  author = {Lee, Seungho and Fujimori, Toshiaki and Nitta, Muneto and Kim, Se Kwon},
  journal = {Phys. Rev. B},
  volume = {112},
  issue = {6},
  pages = {064422},
  numpages = {8},
  year = {2025},
  month = {Aug},
  publisher = {American Physical Society},
  doi = {10.1103/6j6b-sfjh},
  url = {https://link.aps.org/doi/10.1103/6j6b-sfjh}
}

@article{Ivan2020,
  title = {Surface anchoring as a control parameter for stabilizing torons, skyrmions, twisted walls, fingers, and their hybrids in chiral nematics},
  author = {Tai, Jung-Shen B. and Smalyukh, Ivan I.},
  journal = {Phys. Rev. E},
  volume = {101},
  issue = {4},
  pages = {042702},
  numpages = {15},
  year = {2020},
  month = {Apr},
  publisher = {American Physical Society},
  doi = {10.1103/PhysRevE.101.042702},
  url = {https://link.aps.org/doi/10.1103/PhysRevE.101.042702}
}

@Article{Nagaosa2013,
author={Nagaosa, Naoto
and Tokura, Yoshinori},
title={Topological properties and dynamics of magnetic skyrmions},
journal={Nature Nanotechnology},
year={2013},
month={Dec},
day={01},
volume={8},
number={12},
pages={899-911},
issn={1748-3395},
doi={10.1038/nnano.2013.243},
url={https://doi.org/10.1038/nnano.2013.243}
}

@article{Ge2021,
  title = {Observation of Acoustic Skyrmions},
  author = {Ge, Hao and Xu, Xiang-Yuan and Liu, Le and Xu, Rui and Lin, Zhi-Kang and Yu, Si-Yuan and Bao, Ming and Jiang, Jian-Hua and Lu, Ming-Hui and Chen, Yan-Feng},
  journal = {Phys. Rev. Lett.},
  volume = {127},
  issue = {14},
  pages = {144502},
  numpages = {6},
  year = {2021},
  month = {Sep},
  publisher = {American Physical Society},
  doi = {10.1103/PhysRevLett.127.144502},
  url = {https://link.aps.org/doi/10.1103/PhysRevLett.127.144502}
}

@Article{Beneke2024,
author={Beneke, Grischa
and Winkler, Thomas Brian
and Raab, Klaus
and Brems, Maarten A.
and Kammerbauer, Fabian
and Gerhards, Pascal
and Knobloch, Klaus
and Krishnia, Sachin
and Mentink, Johan H.
and Kl{\"a}ui, Mathias},
title={Gesture recognition with {Brownian} reservoir computing using geometrically confined skyrmion dynamics},
journal={Nature Communications},
year={2024},
month={Sep},
day={16},
volume={15},
number={1},
pages={8103},
issn={2041-1723},
doi={10.1038/s41467-024-52345-y},
url={https://doi.org/10.1038/s41467-024-52345-y}
}

@Article{Zhao2023,
author={Zhao, Hanqing
and Malomed, Boris A.
and Smalyukh, Ivan I.},
title={Topological solitonic macromolecules},
journal={Nature Communications},
year={2023},
month={Jul},
day={29},
volume={14},
number={1},
pages={4581},
issn={2041-1723},
doi={10.1038/s41467-023-40335-5},
url={https://doi.org/10.1038/s41467-023-40335-5}
}

@article{Speight2023,
  title = {Symmetries, length scales, magnetic response, and skyrmion chains in nematic superconductors},
  author = {Speight, Martin and Winyard, Thomas and Babaev, Egor},
  journal = {Phys. Rev. B},
  volume = {107},
  issue = {19},
  pages = {195204},
  numpages = {24},
  year = {2023},
  month = {May},
  publisher = {American Physical Society},
  doi = {10.1103/PhysRevB.107.195204},
  url = {https://link.aps.org/doi/10.1103/PhysRevB.107.195204}
}

@article{Speight2020,
  title = {Skyrmions and spin waves in frustrated ferromagnets at low applied magnetic field},
  author = {Speight, Martin and Winyard, Thomas},
  journal = {Phys. Rev. B},
  volume = {101},
  issue = {13},
  pages = {134420},
  numpages = {11},
  year = {2020},
  month = {Apr},
  publisher = {American Physical Society},
  doi = {10.1103/PhysRevB.101.134420},
  url = {https://link.aps.org/doi/10.1103/PhysRevB.101.134420}
}

@article{Benfenati2023,
  title = {Demonstration of $\mathbb{C}{P}^{2}$ skyrmions in three-band superconductors by self-consistent solutions of a {Bogoliubov--de Gennes model}},
  author = {Benfenati, Andrea and Barkman, Mats and Babaev, Egor},
  journal = {Phys. Rev. B},
  volume = {107},
  issue = {9},
  pages = {094503},
  numpages = {10},
  year = {2023},
  month = {Mar},
  publisher = {American Physical Society},
  doi = {10.1103/PhysRevB.107.094503},
  url = {https://link.aps.org/doi/10.1103/PhysRevB.107.094503}
}

@article{Mizushima2002,
  title = {{Mermin-Ho} Vortex in Ferromagnetic Spinor Bose-Einstein Condensates},
  author = {Mizushima, T. and Machida, K. and Kita, T.},
  journal = {Phys. Rev. Lett.},
  volume = {89},
  issue = {3},
  pages = {030401},
  numpages = {4},
  year = {2002},
  month = {Jun},
  publisher = {American Physical Society},
  doi = {10.1103/PhysRevLett.89.030401},
  url = {https://link.aps.org/doi/10.1103/PhysRevLett.89.030401}
}

@article{Mizushima2004,
  title = {Coreless and singular vortex lattices in rotating spinor Bose-Einstein condensates},
  author = {Mizushima, Takeshi and Kobayashi, Naoko and Machida, Kazushige},
  journal = {Phys. Rev. A},
  volume = {70},
  issue = {4},
  pages = {043613},
  numpages = {10},
  year = {2004},
  month = {Oct},
  publisher = {American Physical Society},
  doi = {10.1103/PhysRevA.70.043613},
  url = {https://link.aps.org/doi/10.1103/PhysRevA.70.043613}
}

@article{Gu2014,
  title = {Flexoelectricity and ferroelectric domain wall structures: Phase-field modeling and DFT calculations},
  author = {Gu, Yijia and Li, Menglei and Morozovska, Anna N. and Wang, Yi and Eliseev, Eugene A. and Gopalan, Venkatraman and Chen, Long-Qing},
  journal = {Phys. Rev. B},
  volume = {89},
  issue = {17},
  pages = {174111},
  numpages = {11},
  year = {2014},
  month = {May},
  publisher = {American Physical Society},
  doi = {10.1103/PhysRevB.89.174111},
  url = {https://link.aps.org/doi/10.1103/PhysRevB.89.174111}
}

@article{LeeDH2009,
  title = {Mixed Bloch-N\'eel-Ising character of $180\ifmmode^\circ\else\textdegree\fi{}$ ferroelectric domain walls},
  author = {Lee, Donghwa and Behera, Rakesh K. and Wu, Pingping and Xu, Haixuan and Li, Y. L. and Sinnott, Susan B. and Phillpot, Simon R. and Chen, L. Q. and Gopalan, Venkatraman},
  journal = {Phys. Rev. B},
  volume = {80},
  issue = {6},
  pages = {060102},
  numpages = {4},
  year = {2009},
  month = {Aug},
  publisher = {American Physical Society},
  doi = {10.1103/PhysRevB.80.060102},
  url = {https://link.aps.org/doi/10.1103/PhysRevB.80.060102}
}

@article{Bogdanov1994,
title = {Thermodynamically stable magnetic vortex states in magnetic crystals},
journal = {Journal of Magnetism and Magnetic Materials},
volume = {138},
number = {3},
pages = {255-269},
year = {1994},
issn = {0304-8853},
doi = {https://doi.org/10.1016/0304-8853(94)90046-9},
url = {https://www.sciencedirect.com/science/article/pii/0304885394900469},
author = {A. Bogdanov and A. Hubert}
}

@article{Okubo2012,
  title = {Multiple-$q$ States and the Skyrmion Lattice of the Triangular-Lattice {Heisenberg} Antiferromagnet under Magnetic Fields},
  author = {Okubo, Tsuyoshi and Chung, Sungki and Kawamura, Hikaru},
  journal = {Phys. Rev. Lett.},
  volume = {108},
  issue = {1},
  pages = {017206},
  numpages = {5},
  year = {2012},
  month = {Jan},
  publisher = {American Physical Society},
  doi = {10.1103/PhysRevLett.108.017206},
  url = {https://link.aps.org/doi/10.1103/PhysRevLett.108.017206}
}

@Article{Leonov2015,
author={Leonov, A. O.
and Mostovoy, M.},
title={Multiply periodic states and isolated skyrmions in an anisotropic frustrated magnet},
journal={Nature Communications},
year={2015},
month={Sep},
day={23},
volume={6},
number={1},
pages={8275},
issn={2041-1723},
doi={10.1038/ncomms9275},
url={https://doi.org/10.1038/ncomms9275}
}

@Article{Huh2025,
author={Huh, SeungJung
and Yun, Wooyoung
and Yun, Gabin
and Hwang, Samgyu
and Kwon, Kiryang
and Hur, Junhyeok
and Lee, Seungho
and Takeuchi, Hiromitsu
and Kim, Se Kwon
and Choi, Jae-yoon},
title={Stable singular fractional skyrmion spin texture from the quantum {Kelvin--Helmholtz} instability},
journal={Nature Physics},
year={2025},
month={Sep},
day={01},
volume={21},
number={9},
pages={1398-1403},
issn={1745-2481},
doi={10.1038/s41567-025-02982-x},
url={https://doi.org/10.1038/s41567-025-02982-x}
}

@Article{Fert2013,
author={Fert, Albert
and Cros, Vincent
and Sampaio, Jo{\~a}o},
title={Skyrmions on the track},
journal={Nature Nanotechnology},
year={2013},
month={Mar},
day={01},
volume={8},
number={3},
pages={152-156},
issn={1748-3395},
doi={10.1038/nnano.2013.29},
url={https://doi.org/10.1038/nnano.2013.29}
}

@article{Takeuchi2021_2,
  title = {Quantum Elliptic Vortex in a Nematic-Spin Bose-Einstein Condensate},
  author = {Takeuchi, Hiromitsu},
  journal = {Phys. Rev. Lett.},
  volume = {126},
  issue = {19},
  pages = {195302},
  numpages = {7},
  year = {2021},
  month = {May},
  publisher = {American Physical Society},
  doi = {10.1103/PhysRevLett.126.195302},
  url = {https://link.aps.org/doi/10.1103/PhysRevLett.126.195302}
}

@article{Gerbier2006,
  title = {Resonant control of spin dynamics in ultracold quantum gases by microwave dressing},
  author = {Gerbier, Fabrice and Widera, Artur and F\"olling, Simon and Mandel, Olaf and Bloch, Immanuel},
  journal = {Phys. Rev. A},
  volume = {73},
  issue = {4},
  pages = {041602(R)},
  numpages = {4},
  year = {2006},
  month = {Apr},
  publisher = {American Physical Society},
  doi = {10.1103/PhysRevA.73.041602},
  url = {https://link.aps.org/doi/10.1103/PhysRevA.73.041602}
}

@article{Bookjans2011,
  title = {Quantum Phase Transition in an Antiferromagnetic Spinor Bose-Einstein Condensate},
  author = {Bookjans, E. M. and Vinit, A. and Raman, C.},
  journal = {Phys. Rev. Lett.},
  volume = {107},
  issue = {19},
  pages = {195306},
  numpages = {5},
  year = {2011},
  month = {Nov},
  publisher = {American Physical Society},
  doi = {10.1103/PhysRevLett.107.195306},
  url = {https://link.aps.org/doi/10.1103/PhysRevLett.107.195306}
}

@article{Zhao2014,
  title = {Dynamics in spinor condensates tuned by a microwave dressing field},
  author = {Zhao, L. and Jiang, J. and Tang, T. and Webb, M. and Liu, Y.},
  journal = {Phys. Rev. A},
  volume = {89},
  issue = {2},
  pages = {023608},
  numpages = {5},
  year = {2014},
  month = {Feb},
  publisher = {American Physical Society},
  doi = {10.1103/PhysRevA.89.023608},
  url = {https://link.aps.org/doi/10.1103/PhysRevA.89.023608}
}

@article{Vinit2013,
  title = {Antiferromagnetic Spatial Ordering in a Quenched One-Dimensional Spinor Gas},
  author = {Vinit, A. and Bookjans, E. M. and S\'a de Melo, C. A. R. and Raman, C.},
  journal = {Phys. Rev. Lett.},
  volume = {110},
  issue = {16},
  pages = {165301},
  numpages = {5},
  year = {2013},
  month = {Apr},
  publisher = {American Physical Society},
  doi = {10.1103/PhysRevLett.110.165301},
  url = {https://link.aps.org/doi/10.1103/PhysRevLett.110.165301}
}

@article{Jiang2014,
  title = {Mapping the phase diagram of spinor condensates via adiabatic quantum phase transitions},
  author = {Jiang, J. and Zhao, L. and Webb, M. and Liu, Y.},
  journal = {Phys. Rev. A},
  volume = {90},
  issue = {2},
  pages = {023610},
  numpages = {5},
  year = {2014},
  month = {Aug},
  publisher = {American Physical Society},
  doi = {10.1103/PhysRevA.90.023610},
  url = {https://link.aps.org/doi/10.1103/PhysRevA.90.023610}
}

@article{Seo2015,
  title = {Half-Quantum Vortices in an Antiferromagnetic Spinor Bose-Einstein Condensate},
  author = {Seo, Sang Won and Kang, Seji and Kwon, Woo Jin and Shin, Yong-il},
  journal = {Phys. Rev. Lett.},
  volume = {115},
  issue = {1},
  pages = {015301},
  numpages = {5},
  year = {2015},
  month = {Jul},
  publisher = {American Physical Society},
  doi = {10.1103/PhysRevLett.115.015301},
  url = {https://link.aps.org/doi/10.1103/PhysRevLett.115.015301}
}

@article{Kim2021,
  title = {Emission of Spin-Correlated Matter-Wave Jets from Spinor Bose-Einstein Condensates},
  author = {Kim, Kyungtae and Hur, Junhyeok and Huh, SeungJung and Choi, Soonwon and Choi, Jae-yoon},
  journal = {Phys. Rev. Lett.},
  volume = {127},
  issue = {4},
  pages = {043401},
  numpages = {6},
  year = {2021},
  month = {Jul},
  publisher = {American Physical Society},
  doi = {10.1103/PhysRevLett.127.043401},
  url = {https://link.aps.org/doi/10.1103/PhysRevLett.127.043401}
}

@article{Huh2020,
  title = {Observation of a strongly ferromagnetic spinor Bose-Einstein condensate},
  author = {Huh, SeungJung and Kim, Kyungtae and Kwon, Kiryang and Choi, Jae-yoon},
  journal = {Phys. Rev. Res.},
  volume = {2},
  issue = {3},
  pages = {033471},
  numpages = {9},
  year = {2020},
  month = {Sep},
  publisher = {American Physical Society},
  doi = {10.1103/PhysRevResearch.2.033471},
  url = {https://link.aps.org/doi/10.1103/PhysRevResearch.2.033471}
}

@Article{Huh2024,
author={Huh, SeungJung
and Mukherjee, Koushik
and Kwon, Kiryang
and Seo, Jihoon
and Hur, Junhyeok
and Mistakidis, Simeon I.
and Sadeghpour, H. R.
and Choi, Jae-yoon},
title={Universality class of a spinor Bose--Einstein condensate far from equilibrium},
journal={Nature Physics},
year={2024},
month={Mar},
day={01},
volume={20},
number={3},
pages={402-408},
issn={1745-2481},
doi={10.1038/s41567-023-02339-2},
url={https://doi.org/10.1038/s41567-023-02339-2}
}

@Article{Takeuchi2018,
author = {Takeuchi ,Hiromitsu and Kobayashi ,Michikazu and Kasamatsu ,Kenichi},
title = {Is a Doubly Quantized Vortex Dynamically Unstable in Uniform Superfluids?},
journal = {Journal of the Physical Society of Japan},
volume = {87},
number = {2},
pages = {023601},
year = {2018},
doi = {10.7566/JPSJ.87.023601},
URL = {https://doi.org/10.7566/JPSJ.87.023601}
}

@book{LANDAU1987,
title = {CHAPTER VII - SURFACE PHENOMENA},
booktitle = {Fluid Mechanics},
publisher = {Pergamon},
edition = {2nd},
pages = {238-250},
year = {1987},
isbn = {978-0-08-033933-7},
doi = {https://doi.org/10.1016/B978-0-08-033933-7.50015-5},
url = {https://www.sciencedirect.com/science/article/pii/B9780080339337500155},
author = {L.D. LANDAU and E.M. LIFSHITZ}
}
\end{document}